\documentclass[11pt,a4paper]{article}
\usepackage{jheppub}

\usepackage{graphicx}
\usepackage{amsfonts}
\usepackage{epsfig}
\usepackage{amssymb}
\usepackage{amsmath}

\newcommand{\be}{\begin{eqnarray}}
\newcommand{\ee}{\end{eqnarray}}
\newcommand{\bi}{\begin{itemize}}
\newcommand{\ei}{\end{itemize}}
\newcommand{\Tr}{{\rm Tr}}

\def\lsim{\raise0.3ex\hbox{$<$\kern-0.75em\raise-1.1ex\hbox{$\sim$}}}
\def\gsim{\raise0.3ex\hbox{$>$\kern-0.75em\raise-1.1ex\hbox{$\sim$}}}

\title{The Sphaleron Rate through the Electroweak Cross-over}

\author[a]{Michela D'Onofrio,}
\author[a]{Kari Rummukainen}
\author[b]{and Anders Tranberg}

\affiliation[a]{Department of Physics and Helsinki Institute of Physics,\\ 
  University of Helsinki, P.O.Box 64, FIN-00014 Helsinki, Finland}
\affiliation[b]{Niels Bohr International Academy,
  Blegdamsvej 17, 2100 Copenhagen, Denmark}

\emailAdd{michela.donofrio@helsinki.fi}
\emailAdd{kari.rummukainen@helsinki.fi}
\emailAdd{anders.tranberg@nbi.fi}

\abstract{Using lattice simulations, we measure the sphaleron rate
  in the Standard Model 
  as a function of temperature through the electroweak cross-over,
  for the Higgs masses $m_H=115$ and $m_H=160$\,GeV.
  We pay special attention to the shutting off of the baryon rate as the
  temperature is lowered.
  This quantity enters computations of Baryogenesis via
  Leptogenesis, where non-zero lepton number is converted into non-zero baryon
  number by equilibrium sphaleron transitions. Combining existing numerical methods applicable in the
  symmetric and broken electroweak phases, we find the temperature
  dependence of the sphaleron rate at very high temperature, through
  the electroweak cross-over transition, and deep into the broken
  phase. }


\preprint{HIP, Helsinki}

\begin{document}
\maketitle
\flushbottom

\section{Introduction\label{sec:introduction}}

As a result of a quantum
anomaly, baryon number $B$ and the lepton numbers $L_i$ are not strictly
conserved charges in the Standard Model.   As the Chern-Simons number $N_{\rm CS}$ of the gauge
field evolves in time, because only left-handed fermions are coupled to the SU(2)
gauge field, the baryon and lepton numbers $B$ and $L_i$
change according to the relation (see for instance
\cite{Kuzmin:1985mm,Rubakov:1996vz})
\begin{equation}
\frac{1}{n_G}\left[B(t)-B(0)\right]=L_i(t)-L_i(0)
=N_{\rm CS}(t)-N_{\rm CS}(0)=\frac{1}{32\pi^2}\int_0^t dt\,
\int d^3x\,\Tr\, F^{\mu\nu}\tilde{F}_{\mu\nu},
\label{eq:ncs}
\end{equation}
where $F^{\mu\nu}$ is the field strength tensor of the SU(2) gauge
field and $n_G=3$ is the number of fermion generations in the Standard
Model.

In addition, the electroweak sector of the Standard Model has an
infinite set of ``pure gauge'' degenerate vacua, corresponding to
integer values $N_{\rm CS}$. The question of baryon and lepton number
violation in the Standard Model therefore becomes a question of
whether dynamics allow transitions from one vacuum to another.

At zero temperature transitions occur via instantons
\cite{'tHooft:1976fv}, and the rate is minute, negligible even on
cosmological scales. However at finite temperature, thermal
fluctuations can lead to spontaneous transitions, controlled in
equilibrium by a diffusion (or sphaleron) rate~$\Gamma_{\rm diff}$,
\begin{eqnarray}
\label{eq:diffrate}
\Gamma_{\rm diff}(T)=\lim_{V,\,t\rightarrow \infty}\frac{\langle
  \left[N_{\rm CS}(t)-N_{\rm CS}(0)\right]^2\rangle}{Vt}.  
\end{eqnarray}
At zero temperature, the Standard Model Higgs field acquires a vacuum
expectation value $\langle|\phi|\rangle={v}/{\sqrt{2}}$, 
with $v=246\,$GeV. In contrast, at high temperature 
$
 \langle|\phi|\rangle \approx 0
$,
and there is a ``symmetry breaking'' transition between the two phases
(``symmetric'' and ``broken'').  Note that, because of 
gauge symmetry and Elitzur's theorem, 
$\langle\phi\rangle$ is always zero, whereas
$\langle|\phi|\rangle$  is always non-zero (the value depends on
the renormalization scheme, though).  Thus, we do not
have an exact local order parameter which would vanish in 
the symmetric phase and be non-vanishing in the broken phase.
Indeed, at low Higgs masses ($m_H\, \lsim\, 72$\,GeV), the Standard Model phase 
transition is an actual first-order phase 
transition,
but at experimentally allowed
values for the mass 
the
electroweak transition in the Minimal Standard Model is a continuous
cross-over~\cite{Kajantie:1995kf,Kajantie:1996mn,Rummukainen:1998as,
Budapest}.
Nevertheless, we shall use the labels ``symmetric'' and ``broken'' phases
to describe the states above and below the cross-over temperature.

In the symmetric phase the sphaleron rate is unsuppressed and is proportional
to $T^4$.  In the broken phase the energy barrier between the 
topological sectors grows as $\langle|\phi|\rangle$ increases, and
parametrically the rate is
\begin{equation} 
  \frac{\Gamma_{\rm diff}^{\rm
      brok}(T)}{T^4}=\kappa_{\rm brok}\alpha_w^4e^{-\frac{E_{\rm
        sph}}{T}}, 
\end{equation} 
where $E_{\rm sph}$ is the sphaleron energy (energy barrier) and
$\kappa_{\rm brok}$ is a numerical
coefficient~\cite{Klinkhamer:1984di}.  In the physical cross-over case
it is of interest to know how the rate ``shuts off'' as temperature
decreases, including the precise determination of the rate in the
exponentially suppressed temperature range.  This is important
e.g.~for Leptogenesis scenarios, as discussed in
Sect.~\ref{sec:lepto}.

\subsection{Calculations of the sphaleron rate}
The sphaleron rate has been computed extensively using lattice
simulations. For pure SU(2) gauge theory, the quantum diffusion rate
is approximately recovered in the classical dynamics of a
dimensionally reduced effective theory
\cite{Philipsen:1995sg,Ambjorn:1995xm,Ambjorn:1997jz,Moore:1999fs},
taking into account the proper conversion factors
\cite{Arnold:1996dy,Arnold:1997yb}. The calculation was improved by
including the effect of hard thermal loops
\cite{Moore:1997sn,Bodeker:1999gx}, and the magnitude was finally
settled using the Langevin dynamics of another effective theory
\cite{Bodeker:1998hm,Moore:1998zk}, again using a conversion
prescription~\cite{Moore:1998zk,Arnold:1995bh,Arnold:1999uy}.

The rate in the electroweak theory (i.e.~including the Higgs scalar)
was computed in the symmetric phase at $m_H\simeq 34\,$GeV in
\cite{Ambjorn:1997jz}, and for $m_H\simeq m_W$ in~\cite{Tang:1996qx}
in approximate agreement with the pure-gauge result. The most recent
simulations~\cite{Moore:2000mx} combine all previous methods, extends
the effective theory of~\cite{Bodeker:1998hm} to include the Higgs
field, and finds that at $m_H\simeq 44\,$GeV~\cite{Moore:2000mx}, 
\begin{equation}
  \label{Burnier2.6}
  \frac{\Gamma_{\rm diff}^{\rm sym}}{T^4}=\left(8.24\pm
    0.10\right)\left(\frac{g^2T^2}{m_D^2}\right)
  \left(\ln\left(\frac{m_D^2}{g^2T}\right)+C\right)\alpha_w^5, 
\end{equation} 
with 
\begin{equation} 
  m_D^2=\frac{11}{6}g^2T^2,\qquad C=3.041,\qquad
  \alpha_w=\frac{g^2}{4\pi}.  
\end{equation} 
In the broken phase, the rate is less
well known. The
straightforward numerical methods of~\cite{Ambjorn:1997jz,Tang:1996qx} are not able to
cope with the large suppression of the rate, a problem resolved by
Moore using multicanonical methods~\cite{Moore:1998swa,Moore:1998ge}. In these
papers the low Higgs mass region, where the phase
transition is strongly first order, was considered.  
The latter paper finds that at the transition temperature $T_c$ the 
broken phase rate is
\begin{eqnarray}
  \frac{\Gamma_{\rm diff}^{\rm brok}(T=T_c)}{T_c^4} = \exp\left[-R\left(\frac{m_H}{\textrm{GeV}}\right)\right],
\end{eqnarray} 
with 
\begin{eqnarray} 
  R(49.2)=24.7\pm0.4, \qquad R(44.8)=28.3\pm0.4, \qquad
  R(41.2)=31.2\pm0.6.
\end{eqnarray} 
In this case the strong suppression in the broken phase results in an effectively
instantaneous shut-off of sphaleron processes during the strong
first-order phase transition.  In the cross-over regime, one would
expect a gradual shut-off as the Higgs field expectation value increases.

In~\cite{Shanahan:1998gj} the sphaleron rate was calculated in the
electroweak cross-over region, around $m_H=120\,$GeV, from
high-temperature into the low-temperature phase, using classical
dynamics. Shortly afterwards~\cite{Moore:2000mx}, a similar
calculation was performed at $m_H =$~$130$\,GeV, but using the
effective dynamics of~\cite{Bodeker:1998hm}. In both cases a rather
rapid shut-off of sphaleron diffusion through the cross-over was
reported. However, because no multicanonical methods were used,
the exponentially suppressed tail was not resolved in detail.
The aim of this paper is to improve on these calculations,
and pin down the sphaleron rate through the cross-over, also in the
domain of exponential suppression.

Some time ago~\cite{Burnier:2005hp}, known calculations of the
sphaleron rate were collected and extrapolated to the cross-over
region. The conclusion was that within a range 
\begin{displaymath}
100\, \textrm{GeV} \leq m_H \leq 200\, \textrm{GeV}, 
\end{displaymath}
and for a range of $T$ so that 
\begin{equation}
-\ln\left[\Gamma_{\rm diff}(T)/T^4\right]\simeq 30-50,
\end{equation}
we have
\begin{eqnarray}
  -\ln\left[\frac{\Gamma_{\rm diff}}{T^4}\right] 
   \simeq 
   \sum_{i,j}c_{ij}\left(\frac{m_H-150\,\textrm{GeV}}{10\,\textrm{GeV}}\right)^i
  \left(\frac{T-150\,\textrm{GeV}}{10\,\textrm{GeV}}\right)^j,
\end{eqnarray}
with $c_{00}=39.6$, $c_{10}=3.52$, $c_{01}=-7.09$, $c_{20}=-0.376$,
$c_{11}=0.421$, $c_{02}=0.17$. We will use this result for guidance
and establish its range of validity.

In this work we study the sphaleron rate at Higgs mass
values $m_H=115\,$GeV and $m_H~=~160\,$GeV\footnote{These Higgs
mass values were within the experimentally allowed range
when the simulations were started~\cite{Amsler:2008zzb}.}, 
with special attention
paid at the rate deep in the broken phase.

\subsection{The sphaleron rate and Baryogenesis}

In Electroweak Baryogenesis~\cite{Kuzmin:1985mm,Rubakov:1996vz},
bubble nucleation in a first-order electroweak phase transition
divides space into regions of different electroweak phases. Inside the
bubbles is the broken phase where the sphaleron rate is very
small. Outside the bubbles is the symmetric phase, where the rate is
large. CP-violating interactions between the fermions in the plasma and
the advancing bubble walls generate a net chiral fermion current,
which is then transformed, by sphaleron processes, into a baryon and
lepton asymmetry~\cite{Cohen:1993nk}. A first-order phase transition
is not realized in the Minimal Standard Model, but the Baryogenesis
scenario may be relevant for extensions such as the 2-Higgs Doublet
Model and Supersymmetric Standard Models, in which case a reliable
calculation of the sphaleron rate in both phases is important.

In Baryogenesis via Leptogenesis~\cite{Fukugita:1986hr,Luty:1992un}, a
lepton asymmetry is assumed to originate from some separate process,
represented here by a time-dependent source $f_i(t)$ which may or may
not be active at the electroweak scale. Sphaleron transitions
equilibrate the system, so that the lepton asymmetry is transformed
into net $L_i$ and $B$. Following~\cite{Burnier:2005hp}, the equations
controlling this equilibration read
\begin{eqnarray}
\label{eq:lep1}
\dot{B}(t)&=&-\gamma(t)\left[B(t)+\eta(t)\sum_{i=0}^{n_G}L_i(t)\right],\\
\label{eq:lep2}
\dot{L}_i(t)&=&-\frac{\gamma(t)}{n_G}\left[B(t)+\eta(t)\sum_{i=0}^{n_G}L_i(t)\right]+f_i(t),
\end{eqnarray}
where $T=T(t)$, $\gamma(t)=\gamma[\Gamma_{\rm diff}(T),v(T)]$, $\eta(t)=\eta[v(T)]$ and $n_G=3$. 
$v(T)$ is the expectation value of the Higgs field, taken to be $v(T)\simeq \sqrt{\langle\phi^\dagger\phi\rangle}$.

In this paper, we will calculate $\Gamma_{\rm diff}(T)$ and $v(T)$,
in the Minimal Standard Model at
experimentally allowed Higgs masses, i.e.~in the regime where the
electroweak transition is an equilibrium cross-over.

In Section~\ref{sec:latticemodel} we briefly explain how to treat the
dynamics of the Standard Model at finite temperature as an effective
3-dimensional SU(2)-Higgs theory. Section
\ref{sec:numericalprocedure} we describe the numerical lattice
Monte-Carlo methods employed here, in the high- and low-temperature
regimes, respectively. In Section~\ref{sec:results} we present our
results for $\Gamma_{\rm diff}(T)$ and $v(T)$.
  In Section~\ref{sec:lepto} we solve~(\ref{eq:leptoeq})
for some simple cases to assess the impact of including the correct
sphaleron rate rather than assuming an instantaneous shut-off. We conclude in Section~\ref{sec:conclusion}.

\section{SU(2)-Higgs model on the lattice\label{sec:latticemodel}}

The sphalerons are non-perturbative field configurations and thus
we need to use non-perturbative lattice simulations in order to
calculate the rate reliably.  Furthermore, because of 
the infrared problem in the thermodynamics of the non-Abelian Yang-Mills theory,
the modes with momenta $k \le g^2T$ are non-perturbative.  This means that
the perturbative expansion becomes impossible beyond some loop order 
\cite{Linde:1980ts}.  

\subsection{Dimensional reduction\label{sec:dimred}}

The full four-dimensional Standard Model with chirally coupled
fermions is too unwieldy to simulate on the lattice.  However, for
static (time-independent) thermodynamics we can use the fact that 
the weak coupling constant
is small, and apply perturbation theory only to modes which can be
reliably treated with perturbative methods: that is, to modes with
momentum $k > g^2T$, where $g^2$ is the weak gauge coupling.  
This procedure is called dimensional reduction,
because it results in a three-dimensional effective theory for the
soft ($g^2T$) modes.  The effective theory is purely bosonic, and it
fully includes the essential non-perturbative physics.  The detailed
description how this is performed can be found in
ref.~\cite{Kajantie:1995dw,Farakos:1994kx,Farakos:1994xh}; 
for earlier and related
work, see~\cite{Ginsparg:1980ef,Appelquist:1981vg,Nadkarni:1982kb,Landsman:1989be,Jakovac:1994xg,Braaten:1995cm}.

The perturbative derivation of the effective theory
is based on the hierarchy between the hard ($k\, \gsim\, T$), electric
($k \sim gT$) and magnetic ($k\sim g^2T$) scales on an 
Euclidean finite-temperature path integral.
In the first stage we integrate over the hard scales, obtaining
an effective theory of scales $k\,\lsim\, gT$.  Because
the Matsubara frequencies for the bosonic and fermionic field
modes are $k^{\rm boson}_0 = 2\pi n T$ and 
$k^{\rm fermion}_0 = \pi(2n+1) T$, $n\in Z$
all fermionic modes and non-static ($k_0\ne 0$) bosonic modes are of 
order $T$.  Thus, the effective theory is purely bosonic and three-dimensional.
Concretely, the actual ``integration'' is done by writing down
a general renormalizable effective theory and matching the perturbatively
computed two-, three- and four-point functions in the effective
theory and in the original four-dimensional theory, thus
fixing the parameters of the effective theory.

The effective theory can be further simplified by integrating
over scales $gT$, which gives us the three-dimensional SU(2)
gauge theory coupled to a scalar (Higgs) field:\footnote{%
The hypercharge U(1) and gluon SU(3) gauge fields in principle 
survive the dimensional
reduction.  However, the gluons do not couple to the weak
gauge and Higgs fields and only modify the parameters
in Eq.~\eqref{Oa4} through loop corrections.  
The U(1) field is omitted here because its effect on the transition
is numerically small~\cite{Kajantie:1996mn}.}
\begin{equation}\label{Oa4}
  S = \int d^3x \left( \frac{1}{4} F_{ij}^a F_{ij}^a + \left(D_i \phi \right)^{\dagger} \left(D_i \phi \right) + m_{3}^2 \phi^{\dagger} \phi + \lambda_{3} (\phi^{\dagger} \phi)^2 \right).
\end{equation}
The coupling constants of the theory are $g_3^2 \sim g^2 T$,
$m_3^2$ and $\lambda_3$.  These depend on temperature and the parameters of
the Standard Model; the full expressions are given in
ref.~\cite{Kajantie:1995dw}.  It is customary to use dimensionless
quantities $x$ and $y$ and express the set of the parameters of the
effective theory as
\begin{eqnarray}\label{Kajantie1a}
  g_3^2, \qquad \qquad x = \frac{\lambda_3}{g_3^2}, \qquad \qquad  y = \frac{m_3^2}{g_3^4}.
\end{eqnarray}
Here the dimensionful parameter $g_3^2$ gives the scale and the
physics is completely determined by the values of $x$ and $y$.

\begin{figure}
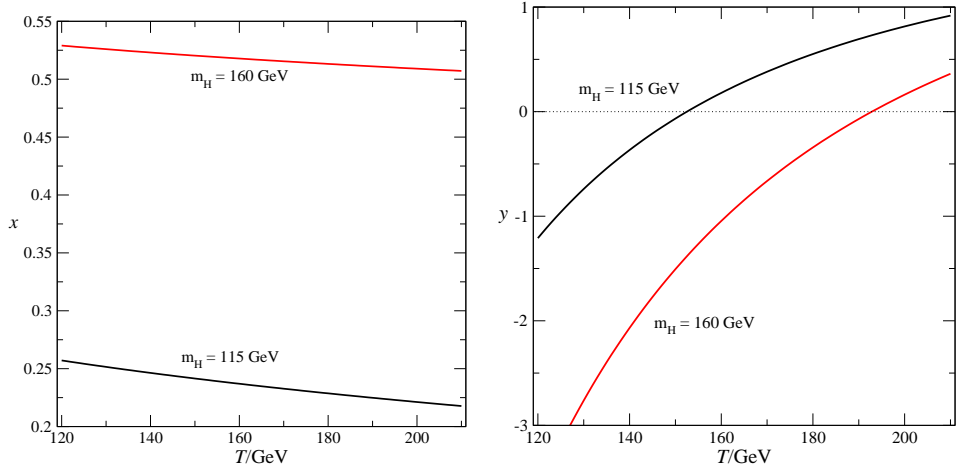

\begin{center}
\includegraphics[width=0.4\textwidth]{xv.eps}~~
\includegraphics[width=0.4\textwidth]{yv.eps}
\end{center}
\caption[a]{The values of $x$ (left) and $y$ (right) for
$m_H=115$ and $160$\,GeV in the temperature range of interest.}
\label{fig:xyvalues}
\end{figure}

The values of $x$ and $y$ for the Higgs masses used in this work,
$m_H=115$\,GeV and $m_H=160$\,GeV, are shown in Fig.~\ref{fig:xyvalues}
over the relevant temperature range.  The other significant Standard
Model parameters which influence the values of $x$ and $y$
are the Fermi coupling $G_F = 1.16639\times10^{-5}$, $m_Z = 91.1876$\,GeV,
$m_W=80.425$\,GeV, $m_t = 174.3$\,GeV and $\alpha_S(m_W) = 0.1187$.

\subsection{Lattice-continuum relations\label{sec:latticecontinuumrelations}}

At this point we have built a continuum 3D theory from the 4D
fundamental one. Now we have to put it on the lattice. 
We use here the common lattice discretization of the 
action~\cite{Kajantie:1995kf}
\begin{eqnarray}
  S_{\rm Lat} 
  &=& \beta_{G} \sum_x \sum_{i<j} \left(1- \frac{1}{2}\textrm{Tr}\left[P_{ij}\right]\right)-\beta_H \sum_x \sum_{i} \frac{1}{2} \textrm{Tr}\left[\Phi^{\dagger}(x)U_i(x)\Phi(x+\hat i)\right] + \notag \\
  &+& \sum_x \frac{1}{2}
  \textrm{Tr}\left[\Phi^{\dagger}(x)\Phi(x)\right]+ \beta_R \sum_x
 \left[\frac{1}{2}\textrm{Tr}\left[\Phi^{\dagger}(x)\Phi(x)\right]-1\right]^2, \label{KLRS2.4}
\end{eqnarray}
which is constructed only with gauge-invariant terms. Here $U_i(x)$ 
is the SU(2) gauge link variable,  $P_{ij}$ is the standard $ij$-plane plaquette
constructed from the link variables, and the lattice Higgs field is 
naively scaled from the continuum field with
$(1/8)\beta_G \beta_H \Phi^\dagger\Phi = \phi^\dagger\phi/g_3^2$.
The parameters
$\beta_G$, $\beta_H$ and $\beta_R$ are related to the parameters
$g_3^2a$, $x$ and $y$ by 
\begin{eqnarray}
  \beta_G &=& \frac{4}{g_3^2 a}, \label{KLRS2.5} \\
  x &=& \frac{1}{4} \lambda_3 a \beta_G = \frac{\beta_R \beta_G}{\beta_H^2}, \label{KLRS2.6}\\
  y &=& \frac{\beta_G^2}{8}\left(\frac{1}{\beta_H}-3-\frac{2x\beta_H}{\beta_G}\right)+ \frac{3\Sigma \beta_G}{32 \pi}(1+4x)+ \frac{1}{16 \pi^2} \cdot \notag \\
  && \cdot \left[\left(\frac{51}{16}+9x-12x^2\right)\left(\ln\frac{3 \beta_G}{2}+\zeta\right)+5.0+5.2x\right]. \label{KLRS2.7}
\end{eqnarray}
Here $\Sigma = 3.1759115$ and $\zeta = 0.08849$ and the other numerical constants appearing in~(\ref{KLRS2.7}) are specific for the SU(2) $+$ Higgs model and are
computed in~\cite{Laine:1995ag}.  

We note here that because the action is superrenormalizable, only the
mass term ($y$) gets renormalized.  The continuum limit is taken using
equations~\eqref{KLRS2.5}--\eqref{KLRS2.7} by letting
$\beta_G\rightarrow \infty$ while keeping $x$, $y$ and $g_3^2$
constant.  The counterterms in Eq.~\eqref{KLRS2.7} remove the linear
and logarithmic divergences in the lattice spacing $a$.  

We actually implement the following simple improvement to equations~\eqref{KLRS2.5}--\eqref{KLRS2.7}, which cancels part of the
$O(a)$ corrections~\cite{Moore:1997np}: 
in the lattice action~\eqref{KLRS2.4} and equations~\eqref{KLRS2.6}--\eqref{KLRS2.7} we substitute 
\begin{equation}
  \beta_G \rightarrow \beta_{G,\rm improved} = 4/(g_3^2 a) + 0.6674\,,
\end{equation}
which makes the gauge part of the action $O(a)$ accurate.
In fact,~\eqref{KLRS2.6} can also be $O(a)$ improved~\cite{Moore:1997np}, but
since the full $O(a)$ correction to~\eqref{KLRS2.7} is not known we do not
implement it here.  In what follows we consistently use the identity 
$\beta_G = 4/(g_3^2a)$.

The gauge invariant Higgs field expectation value $\langle \Phi^\dagger\Phi \rangle$ has linear and logarithmic UV divergences.  Subtracting the divergences from the lattice expectation value, we obtain the continuum quantity $\langle \phi^{\dagger}\phi \rangle$ as
\cite{Kajantie:1995kf} using
\begin{equation}
\label{KLRS4.2}
\frac{\langle \phi^{\dagger}\phi \rangle}{g_3^2}= 
\frac{1}{8} \beta_G \beta_H \left( \langle \Phi^\dagger\Phi \rangle - \frac{\Sigma}{\pi \beta_H} \right)-\frac{3}{(4 \pi)^2} \left(\log \frac{3 \beta_G g_3^2}{2 g_3^2}+\zeta+ \frac{1}{4}\Sigma^2 -\delta \right)+ O \left(\frac{1}{\beta_G}\right)
\end{equation}
where $\zeta+ \frac{1}{4}\Sigma^2 -\delta \approx 0.6678$.

\subsection{Real-time evolution}

The effective theory in Eq.~\eqref{Oa4} is well understood and has
been very successfully used in studies of static thermodynamical quantities
of hot electroweak physics.  As such, it does not describe 
dynamical phenomena, which include sphaleron transitions.  
It is possible to take the theory in Eq.~\eqref{Oa4} and use
the classical equations of motion to describe the time evolution
of the fields, as was done in the early studies of the 
sphaleron rate~\cite{Ambjorn:1997jz,Tang:1996qx,Moore:1998swa}.
However, it has been shown that the classical theory contains 
divergent UV contributions to the gauge field dynamics, and the
results are cut-off dependent~\cite{classtheorysick}.  Hence,
technically the infrared gauge field dynamics of the classical
theory do not exist.  The physical origin of the problem 
is that the Landau damping of the transverse gauge fields in the
classical theory is UV divergent, and the theory does not have a 
physical continuum limit.

These problems can be ameliorated by studying classical theory with hard thermal loop (HTL) effects included~\cite{HTL_papers}.  This leads to complicated and expensive numerical implementations~\cite{Moore:1997sn,Bodeker:1999gx}.  However, as first demonstrated
by B\"odeker~\cite{Bodeker:1998hm}, the physical damping makes the dynamics of the infrared gauge field modes (modes with $k\,\gsim\, g^2 T$) to be fully overdamped. Then, at leading order in $1/\ln(1/g)$ the evolution of these modes is described with simple Langevin dynamics (in $A_0=0$ gauge, and identifying $H/T = S$ in~\eqref{Oa4}) 
\cite{Arnold:1999uy}:
\begin{equation}
 \partial_t A_i = - \sigma_{\rm el}^{-1} \frac{\partial H}{\partial A_i}
  + \xi_i^a,
\end{equation}
where $\sigma_{\rm el}$ is the non-Abelian ``color'' conductivity,
\begin{equation}
  \sigma_{\rm el}^{-1} = \frac{3\gamma}{m_D^2},
~~~~
  \gamma = \frac{2g^2T}{4\pi}\left(\ln\frac{m_D}{\gamma} + 3.041\right),
\end{equation}
and again $m_D^2 = (11/6)g^2 T^2$ in the standard model.
$\xi$ is a random Gaussian noise with
\begin{equation}
  \langle \xi_i^a({\bf x},t) \xi_j^b({\bf x'},t')\rangle = 
  2\sigma_{\rm el} T\delta_{ij}\delta^{ab}\delta({\bf x} - {\bf x'})\delta(t-t').
\end{equation}
The Higgs field has parametrically much less damping.  Hence, it can also be evolved with a Langevin equation, but with a much faster rate of evolution. To this accuracy we can take it to be infinitely fast in comparison with the gauge field evolution~\cite{Moore:2000mx}. 
Iterating, we can solve for $\gamma = 0.66361688 \ g^2 T$.

In principle the Langevin evolution is straightforward to implement on the lattice.  However, it is unnecessarily slow: we can substitute it with any dissipative update, as long as the relation between the evolution rates is known.  Thus, it is much more efficient to use random-order heat-bath update algorithm for the SU(2) gauge fields~\cite{Moore:1998zk,Moore:2000jw}.
Now $n$ full heat-bath update sweeps through the lattice correspond
to the real-time step 
\begin{equation}
\label{Moorep28}
\Delta t = \frac14 \sigma_{\rm el} \,a^2 n\,.
\end{equation} 
We note that this relation is valid for ``unimproved'' single-plaquette
Wilson gauge and random-order heat bath; for other choices the relation
would differ.  The Higgs field is updated with a mixture of heat bath and overrelaxation much more frequently than the gauge field~\cite{Moore:2000jw}.

\section{Measuring the sphaleron rate\label{sec:numericalprocedure}}

\begin{figure}
\centerline{
\includegraphics[width=0.55\textwidth]{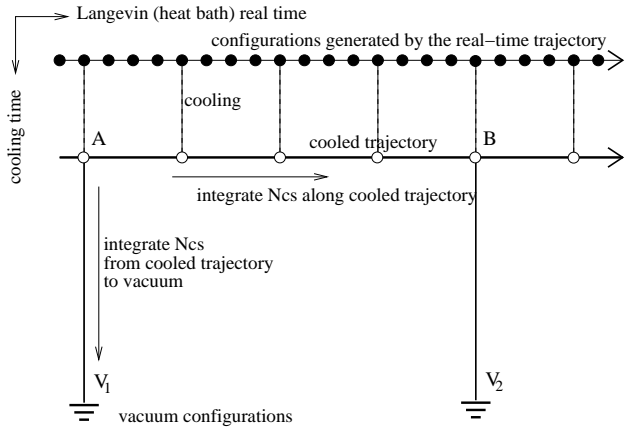}
}
\caption[a]{Measurement of the Chern-Simons number evolution~\cite{Moore:1998swa}. The solid circles show the configurations generated by the real-time evolution using the Langevin/heat-bath method.
At fixed intervals, the configurations are cooled by the same amount in order to construct a cooled trajectory, where the UV noise is almost completely eliminated, allowing to calculate $\delta N_{\rm CS}$ from~\eqref{HTL6.3}. The cooling from vacuum to vacuum works as a test for residual errors: $\delta N_{\rm CS}$ must then be close to an integer, the deviations from which are subtracted, thus avoiding the accumulation of errors.\label{fig:cooling}}
\end{figure}

The evolution of the Chern-Simons number $N_{\rm CS}$ over a time interval
$(t_0, t)$ can be defined using lattice electric and magnetic fields:
\begin{eqnarray}
\label{HTL6.3}
\delta N_{\rm CS}(t) \equiv N_{\rm CS}(t)-N_{\rm CS}(t_0)
=\frac{g^2}{8 \pi^2}\int_{t_0}^t dt' \int d^3x E_i^a B_i^a.
\end{eqnarray}
Unfortunately the topology on the lattice is not well defined, and
using naive lattice scale $E$ and $B$ fields the right-hand side of 
equation~\eqref{HTL6.3} contains ultraviolet noise.  This gives unphysical diffusion not connected with the sphaleron rate.   
The method of {\em calibrated cooling}~\cite{Moore:1998swa} (see also~\cite{Ambjorn:1997jz}),
offers a way out of the problem. It is based on the fact
that at small enough lattice spacing sphalerons are large in lattice units, with a dominant length scale of order $1/(g^2 T)$.\footnote{%
In ref.~\cite{Moore:1999fs} the size of the sphaleron was estimated to be
of order $5/(g^2 T)$ in the pure SU(2) gauge theory.}
By applying a pre-determined amount of cooling (Langevin evolution without the noise) to the lattice gauge fields, the
ultraviolet noise is eliminated, without compromising the long-distance topology of the configuration.  At this point it is possible to evaluate the
integral in~(\ref{HTL6.3}) with only small errors.  Cumulative residual
errors are eliminated by periodically cooling all the way down to a vacuum
configuration and correcting for deviation from integer values of
$\delta N_{\rm CS}$ between two vacua.  This is schematically described in Figure~\ref{fig:cooling}.  By adjusting the cooling parameters so that these deviations from integers always remain much smaller than unity,
we also ensure that the cooling is sufficient to keep the measurement topological.   

Cooling the original gauge fields close to the vacuum is computationally very expensive.  The procedure is dramatically accelerated by blocking the lattice gauge fields by a factor of two after the UV noise has been sufficiently eliminated; this is repeated a couple of times until a minimum lattice size has been reached.  For details, we refer to~\cite{Moore:1998swa}.

\subsection{Sphaleron rate in the symmetric phase}
\label{sec:symmetric}

We calculate the sphaleron rate using two different, and
complementary, methods. We start at high temperatures, above the
cross-over, and go through the whole cross-over range into the ``broken
phase'' by uniformly decreasing the temperature. At high temperatures
we use standard canonical Monte Carlo sampling. As the potential
barrier between consecutive Chern-Simons numbers is low, the
probability distribution over Chern-Simons number is approximately
flat. As we decrease the temperature, the sphaleron rate becomes
exponentially suppressed and the canonical real-time method is too inefficient
to resolve the slow rate.  At this point we begin using
multicanonical simulations.

\begin{figure}
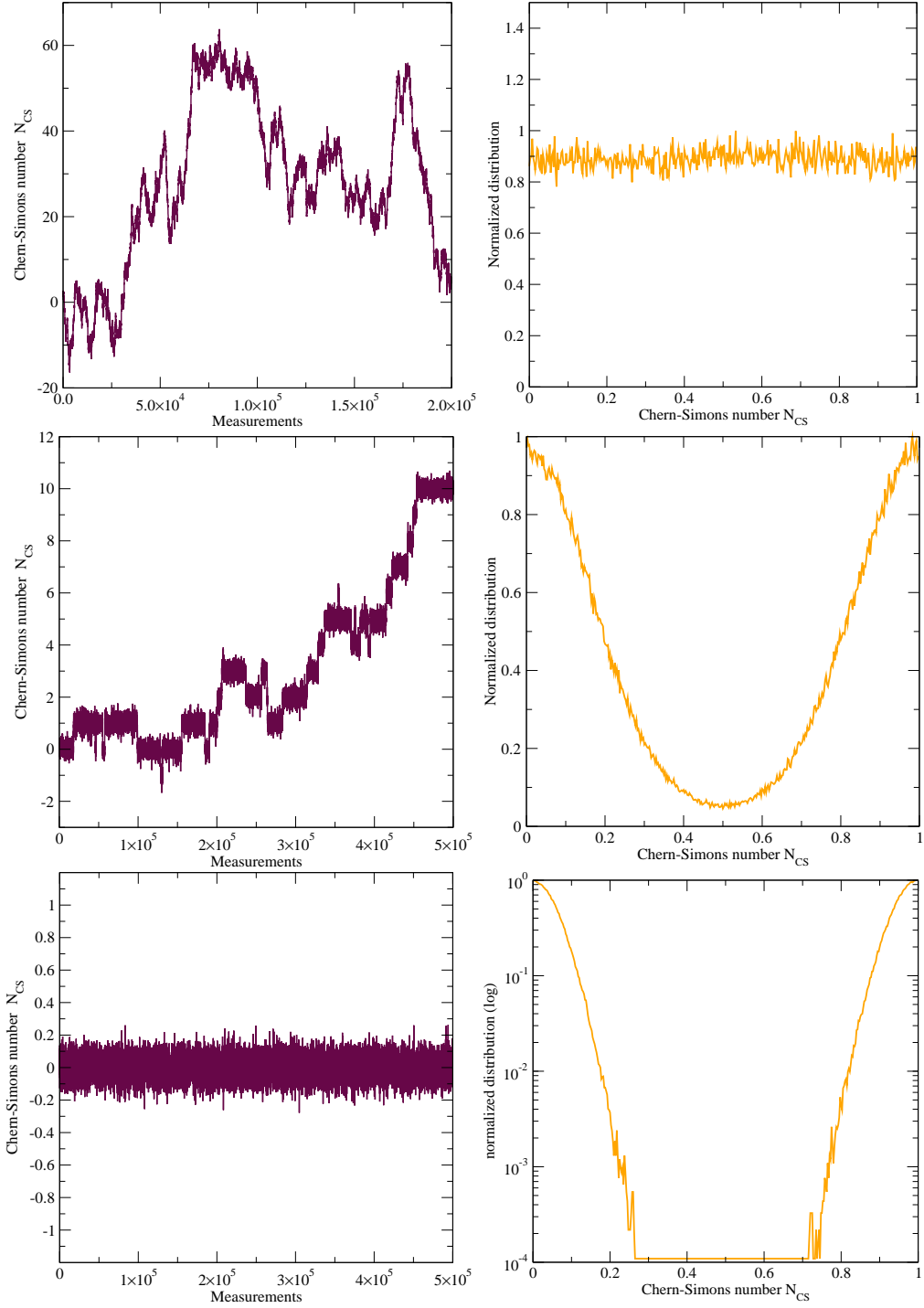

\includegraphics[height=0.41\textwidth]{ncs_m113t152_nt.eps}
\includegraphics[height=0.41\textwidth]{p_m113t152_nt.eps} \\
\includegraphics[height=0.42\textwidth]{./ncs_m113t145_nt} 
\includegraphics[height=0.42\textwidth]{./p_m113t145_nt}\\
\hspace*{-2.5mm}
\includegraphics[height=0.415\textwidth]{./ncs_m113t140_nt}
\includegraphics[height=0.415\textwidth]{./plog_m113t140_nt}
\caption[a]{$N_{\rm CS}$ from a heat-bath trajectory (left), and
the resulting probability distribution (right), folded into the interval $[0,1]$, at $m_H =115\,$GeV and $T =152$ (top), $145$ (middle) and $140$\,GeV (bottom).  At high temperature, in the symmetric phase, the 
sphaleron transitions are unsuppressed, whereas at low $T$ 
the transitions are so strongly suppressed that they do not
happen in canonical simulations.}
\label{fig:ncs_hb}
\end{figure}

As an example, the evolution of the Chern-Simons number
at $m_H=115$\,GeV and with $T=152$\,GeV (symmetric phase), $145$\,GeV (cross-over region) and $140$\,GeV (broken phase) is shown in Figure~\ref{fig:ncs_hb}.  In the symmetric phase the transitions are unsuppressed and it is straightforward to measure the diffusion rate.  Around the cross-over temperature the probability distribution of $\Delta N_{\rm CS}$ becomes peaked around integer values and the transitions between these values become rapidly more suppressed.  Finally, deep in the broken phase the
rate goes down until we are not able to measure it with the real-time evolution method.

\subsection{Sphaleron rate in the broken phase: multicanonical method}

\begin{figure}[ht]
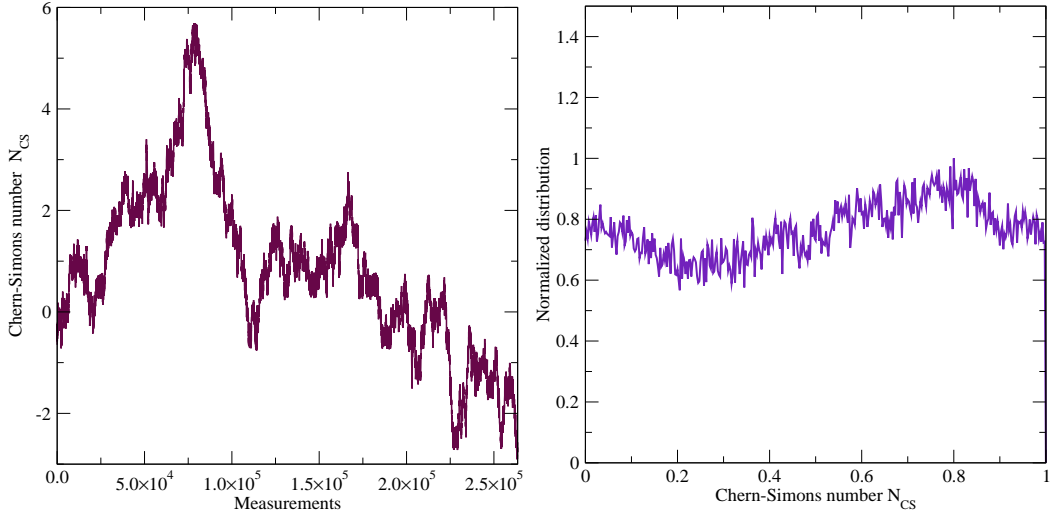

\label{fig:ncs_broken_multi} 
\begin{center}
  \includegraphics[width=0.45\textwidth]{./ncs_m113t140mc_nt} 
  \includegraphics[width=0.45\textwidth]{./pmul_m113t140mc_nt} 
\end{center}
\caption[a]{A heat-bath trajectory for $N_{\rm CS}$, still for $m_H =$ 115\,GeV and $T =$ 140\, but now with multicanonical simulations (left), and the corresponding multicanonical probability distribution $P_{\textrm{muca}}$ (right).}
\label{fig:muca}
\end{figure}

\begin{figure}[ht]
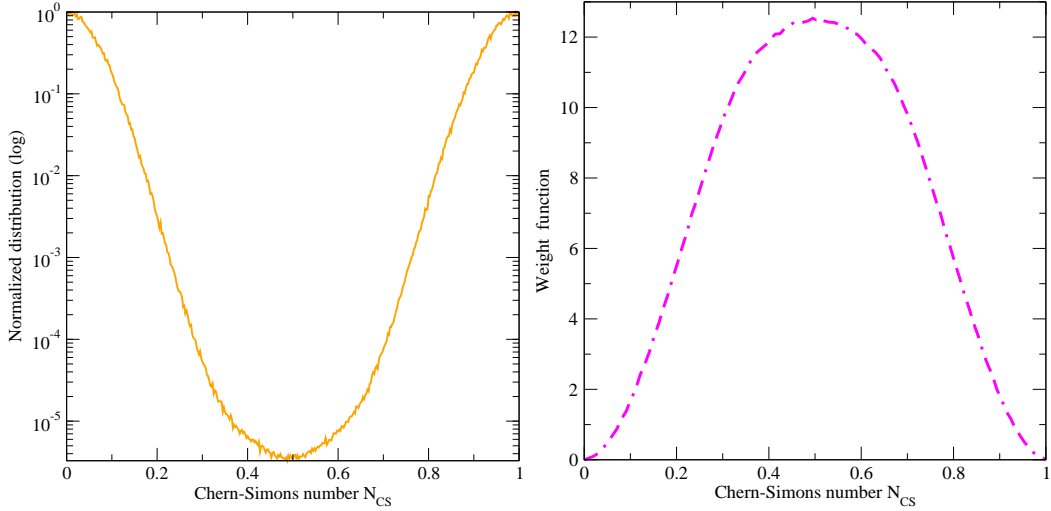

\centerline{
  \includegraphics[width=0.45\textwidth]{./plog_m113t140mc_nt} 
  \includegraphics[width=0.45\textwidth]{./weight_m113t140nt}
}
\caption[a]{The physical distribution $P_{\textrm{can}}$ (left), after reweighting the result in figure~\ref{fig:ncs_broken_multi} with the multicanonical weight function (right).}
\label{fig:weight}
\end{figure}

At low temperatures where the sphaleron rate is strongly suppressed,
this can be measured using a multicanonical method similar to the one
used in~\cite{Moore:1998swa}.  The calculation consists of two stages:\\
i) the measurement of the potential barrier (probabilistic suppression) between two integer vacua, and \\
ii) the calculation of the rate of the tunneling through the top of the potential barrier.  

Let us first look at the measurement of the potential barrier.  This is the multicanonical stage of the computation.  As is usually done in a multicanonical context, instead of sampling configurations with the
canonical weight
\begin{eqnarray}\label{can_p}
  P_{\textrm{can}}(U,\Phi) \propto e^{-S(U,\Phi)},
\end{eqnarray}
we compensate for the strong suppression by adding a carefully-chosen
weight function $W(N_{\rm CS})$, which is a function of the order parameter, in this case the Chern-Simons number.  The configurations $\{U,\Phi\}$ are now sampled with the probability density
\begin{eqnarray}\label{multi_p}
  P_{\textrm{muca}}(U,\Phi) \propto e^{-S(U,\Phi) + W(N_{\rm CS}[U])}.
\end{eqnarray}
Defining the physical (canonical) probability distribution of the Chern-Simons
number
\begin{equation}
  p_{\textrm{can}}(N'_{\rm CS}) = 
  \int {\rm d}U {\rm d}\Phi \, 
  P_{\textrm{can}}[U,\Phi] \delta(N'_{\rm CS}-N_{\rm CS}[U]),
\end{equation}
it is clear that the corresponding multicanonical distribution is
\begin{equation}
  p_{\textrm{muca}}(N_{\rm CS}) = 
  p_{\textrm{can}}(N_{\rm CS}) e^{W(N_{\rm CS})}.
  \label{pmuca}
\end{equation}
Thus, the probability suppression in multicanonical simulations vanishes if we choose $W(N) = - \ln p_{\textrm{can}}(N) + \mbox{const}$.
This is not a particularly useful result because we do not know the canonical distribution a priori; indeed, that is the quantity we set out to compute with the multicanonical method.  

However, it is possible to calculate a good enough approximation for $W$ by using an automatic iterative ``self-learning'' procedure.  Here we follow the method presented in ref.~\cite{Moore:2000jw}, with the obvious modification of using a different order parameter.  Essentially, during the learning stage the weight function is continuously modified in order to maximize the flatness of the total distribution of the Chern-Simons number.  When the iteration has sufficiently converged, the resulting weight function $W$ is then used in a production run.  

The physical (canonical) Chern-Simons probability distribution is now obtained from the measured multicanonical distribution using equation~\eqref{pmuca}.
An example of the multicanonical evolution and the resulting probability distribution is shown in Figure~\ref{fig:muca}, for $m_H=115$\,GeV and $T=140$\,GeV; the same parameters as the lowest temperature in Figure~\ref{fig:ncs_hb}.  We recall that in this figure the ``evolution'' cannot be interpreted as a physical evolution in real time. As we can observe, the distribution of the Chern-Simons number is now almost flat.  The resulting canonical (physical) probability distribution is shown in Figure~\ref{fig:weight}, together with the weight function $W(N_{\rm CS})$ used.

We obtain an estimate for the physical expectation value for a general observable $A$ from multicanonical simulation from
\begin{equation}
  \langle A \rangle = 
  \frac{\sum_{i} A_i e^{-W_i}}
       {\sum_{i}  e^{-W_i}},
\end{equation}
where the sums go over the configurations $\{U,\Phi\}_i$ obtained from the simulation and on which the measurements are performed, $W_i = W(N_{\rm CS}[\{U\}_i])$ and $A_i = A(\{U,\Phi\}_i)$.

A couple of comments on the practical implementation of the multicanonical sampling using the Chern-Simons number are in order.  Because now the Chern-Simons number enters in the sampling weight for the configurations, it has to be evaluated in an unbiased manner; that is, the Chern-Simons number for a given gauge field configuration has to be a unique value (modulo 1), independent of the ``history'' of the configuration.  The calibrated cooling described above has a small residual component which depends on the trajectory of the configurations.  To ensure unbiased sampling, we do not use the cooled trajectory method but cool down to the vacuum at every measurement of $N_{\rm CS}$.  More precisely, we first cool the gauge field a pre-determined amount (in most cases the cooling time is $\sim 8.4\,a^2$) in order to reduce the UV noise, and then start integrating ${\bf E}\cdot{\bf B}$ while cooling to vacuum.  The frequent cooling now completely dominates the CPU-time budget of the simulation.

The multicanonical probabilistic weight is implemented as an accept/reject step as follows: i) start with configuration $A$, with weight function $W_A$; ii) perform one heat-bath sweep through the lattice, producing provisional new configuration $B$; iii) measure $N_{\rm CS}(B)$ as described above, obtaining $W_B$.  iv) The new configuration is accepted with the probability
\begin{equation}
  p_{\rm accept}(A\rightarrow B) = \left\{ 
      \begin{array}{ll}
        1 & \mbox{if~}W_A \le W_B \\
        e^{W_B - W_A} &\mbox {if~} W_A > W_B
      \end{array}
      \right. .
\end{equation}
If the update is rejected, we start again at point i) with configuration $A$.
The acceptance rate was around 50\% at the lowest temperatures used, and increasing at higher temperatures.

Obviously, the measured value of $N_{\rm CS}$ depends on the amount of cooling applied before the measurement.  Thus, the obtained probability distribution $p(N_{\rm CS})$ is
also cooling dependent.  However, this is completely cancelled by the dynamical rate measurement described in Section~\ref{sec:methodBro}, so that the final
rate is independent of the amount of cooling.  Nevertheless, the right amount
of cooling must be judiciously chosen for efficiency: insufficient cooling gives too noisy observables, whereas too much cooling takes one too far ``downhill'' from the original configuration towards the vacuum.  In both cases the measured $N_{\rm CS}$ is too far from the ``true'' value, and the result is that we do not observe a random walk for $N_{\rm CS}$.  The situation becomes worse at large volumes and coarse lattices.

\subsection{Sphaleron rate in the broken phase: dynamical prefactor\label{sec:methodBro}}

The multicanonical procedure described above gave us the probability distribution of the Chern-Simons number in the broken phase.  We can now measure the tunneling rate following refs.~\cite{Moore:1998swa,Moore:2000jw}:
\begin{enumerate}
\item Let us assume that we have done the multicanonical
  simulations and obtained the canonical (physical) probability distribution
  of the Chern-Simons number \ $p_{\rm phys.}(N_{\rm CS})$. 

\item We choose a narrow interval $1/2 - \epsilon/2 \le N_{\rm CS} \le
  1/2 + \epsilon/2$ around the point that separates vacuum $N_{\rm CS}
  = 0$ from the vacuum $N_{\rm CS}=1$. The relative probability of
  finding a configuration here is
  \begin{equation}
    P(|N_{\rm CS} - 1/2| < \epsilon/2) =
    \int_{1/2-\epsilon/2}^{1/2+\epsilon/2} dN p_{\rm phys}(N).
  \end{equation}
  This is where we need multicanonical
  methods, as the probability of being on top of the
  barrier is extremely small, and to get a reliable estimate would
  take an impractically long time with canonical sampling. 

\begin{figure}
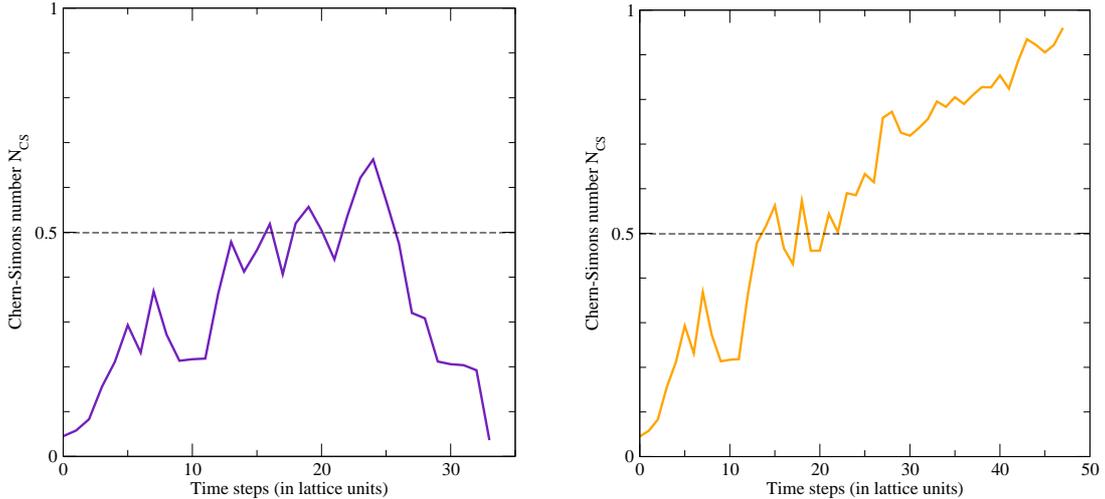
 
\centerline{
\includegraphics[width=0.445\textwidth]{./trajec2.eps}
\hspace*{0.6cm}
\includegraphics[width=0.45\textwidth]{./trajec1.eps}
}
\caption[a]{Two real-time trajectories starting from the same
  configuration. The final and initial configuration can either be the same (left) or different (right).
The trajectories cross the central value of our order parameter $N_{\rm CS}=1/2$ several times, a fact we compensate for through the dynamical prefactor~(\ref{dynpref}).}
\label{fig:traj}
\end{figure}

\item Let us now take a random configuration from the canonical
  distribution but with the constraint $1/2 - \epsilon/2 < N_{\rm CS}
  < 1/2 + \epsilon/2$; i.e.~near the top of the potential barrier. Starting from this
  configuration, we now generate two real-time trajectories using the heat-bath dynamics, as described in Section~\ref{sec:symmetric}.  
  The trajectories are evolved until the
  Chern-Simons number falls near a vacuum value.  Interpreting one of
  the trajectories as evolving backwards in time, we can glue the
  trajectories together at the starting point and obtain a
  vacuum-to-vacuum trajectory.  The trajectory can either return to
  the starting vacuum or be a genuine tunneling trajectory, see
  Figure~\ref{fig:traj}.  Only the latter-type trajectories contribute to the
  sphaleron rate.

\item We can obtain the tunneling rate by measuring 
  $|\Delta N_{\rm CS} / \Delta t|$ from the trajectories at the moment they cross the
  value $N_{\rm CS}=1/2$.  Here $\Delta t$ is the time interval between successive measurements, and $\Delta N_{\rm CS}$ the change in Chern-Simons number.  This characterizes the probability flux thorough 
  the top of the barrier.   We obtain the physical time difference from
  the relation between the heat-bath ``time'' and physical time, equation~\eqref{Moorep28}.

\item
  If the tunneling trajectories would go straight across the top, the ingredients above would be sufficient to calculate the total rate.  However, typically the trajectories ``random walk'' near the top of the barrier and can cross the value $N_{\rm CS}=1/2$ several times.  Because the trajectories were chosen starting from a set of configurations near the top of the barrier, this leads to overcounting: the evolution could be started at any point the $N_{\rm CS}=1/2$ limit is crossed.  This can be compensated by calculating a dynamical prefactor 
  \begin{eqnarray}\label{dynpref}
  \textrm{d} = \frac{1}{N_{\rm traj}} \sum_{\rm traj} \frac{\delta_{\rm tunnel}}{\# \ \textrm{crossings}},
\end{eqnarray}
where the sum goes over the ensemble of trajectories, $N_{\rm traj}$ is the number of trajectories, $\delta_{\rm tunnel}$ is 0 if the trajectory does not lead to a change of the vacuum and 1 if it does, and (\# crossings) is the number of times the trajectory crosses $N_{\rm CS}=1/2$.  
\end{enumerate}
With these ingredients, the sphaleron rate now becomes
\begin{eqnarray}
\Gamma = 
\frac{P(|N_{\rm CS} - 1/2| < \epsilon/2) }{\epsilon}  
\left\langle  \left| \frac{\Delta N_{\rm CS}}{\Delta t} \right| \right\rangle  \textrm{d}.
\label{Gamma_mc}
\end{eqnarray}
We note that the result is independent of $\epsilon$ as long as $\epsilon \ll 1$.
It is also independent of the frequency $\Delta t$ with which the Chern-Simons number is measured:  if we decrease the measurement interval, the trajectories become more jagged due to the random-walk nature of the heat-bath updates.  This will increase the number of the crossings of the value $N_{\rm CS}~=~1/2$ and hence decrease $\textrm{d}$.  However, the latter is completely compensated by a corresponding increase in $\langle | \Delta N_{\rm CS}/\Delta t|\rangle$.  If the measurement interval $\Delta t$ is small enough, random walk arguments imply $\textrm{d} \propto (\Delta t)^{1/2}$ and $\langle | \Delta N_{\rm CS}/\Delta t|\rangle \propto (\Delta t)^{-1/2}$.  This is corroborated by the numerical data.  Thus, equation~\eqref{Gamma_mc} has a well-defined continuum limit.

\section{Results\label{sec:results}}

We concentrate on two physically significant observables, the sphaleron rate and the Higgs field expectation value as functions of temperature at $m_H=115\,$GeV and $m_H=160$\,GeV.  For both quantities we check for the
finite volume and finite lattice spacing effects.

\subsection{Higgs field $v(T)$\label{sec:resultvev}}

\begin{figure}
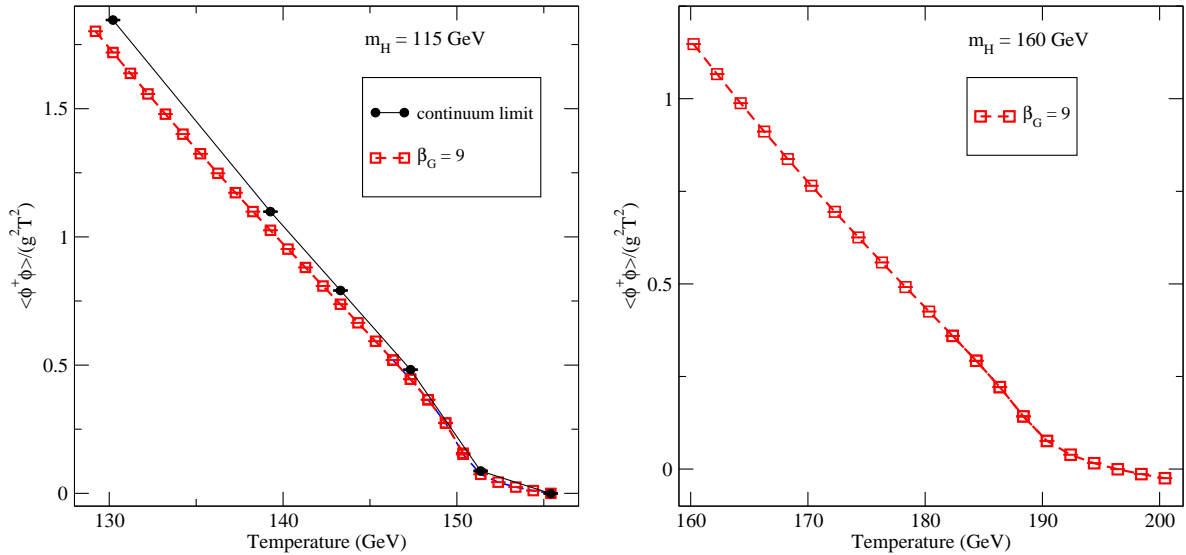

\centerline{%
\includegraphics[width=0.5\textwidth]{phi115naive.eps}%
~~
\includegraphics[width=0.5\textwidth]{phi160naive.eps}
}
\caption[a]{The Higgs field expectation value for $m_H =115\,$GeV (left) and
$160$\,GeV (right). Squares are for $\beta_G = 9$, volume $(L/a)^3 = 32^3$.  For 
$m_H=115$\,GeV we have performed the continuum limit extrapolation at 
selected temperatures using 
a range of lattice spacings $\beta_G= 4/(g_3^2 a) = 6\ldots 12$ and extrapolating linearly
to continuum.  We observe that the $\beta_G=9$ result deviates less
than 8\% from the continuum limit in the range of temperatures studied.
The lines are to guide the eye. }
\label{fig:phi2}
\end{figure}

The gauge invariant Higgs condensate $\langle \phi^\dagger \phi\rangle$
is a direct probe of the phase transition or cross-over.  At high temperatures
it is close to zero, and at low temperatures it acquires an expectation value which
grows as the temperature decreases.   Because of the additive renormalisation, Eq.~\eqref{KLRS4.2}, the symmetric-phase value can become negative.
The results obtained at $\beta_G=9$, lattice size $32^3$, are shown in Fig.~\ref{fig:phi2}.  Note that we display $\langle\phi^2\rangle$ in units of $g^2T^2$, and goes to infinity as temperature goes to zero.  

At high temperature in the symmetric phase, $\langle\Phi^\dagger\Phi\rangle$ is close, but not quite identical, to zero. As the temperature is lowered, we enter the cross-over region where $\langle\Phi^\dagger\Phi\rangle$ grows rapidly.  At smaller Higgs masses ($m_H~\lsim~72$\,GeV), this rapid growth becomes a discontinuous jump, indicating a first-order transition~\cite{Kajantie:1996mn}.  Below the cross-over region, $\langle\Phi^\dagger\Phi\rangle/(g^2T^2)$ settles to an almost linear increase.  

\begin{figure}
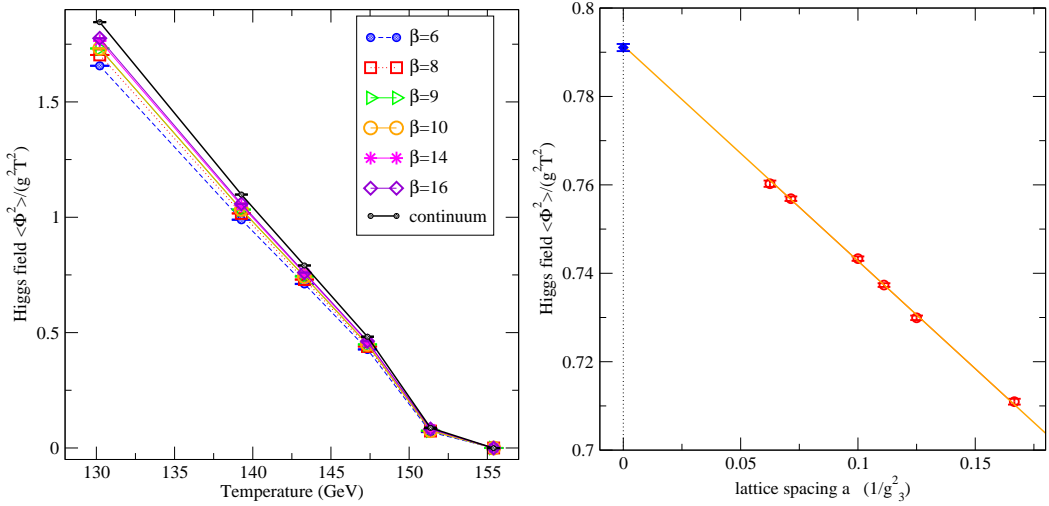

\begin{center}
\includegraphics[width=0.45\textwidth]{phi_fixedratio_naive.eps}
\includegraphics[width=0.45\textwidth]{aphi2t143naive.eps}
\end{center}
\caption[a]{Left: the Higgs field calculated for $m_H =115\,$GeV and
  several temperatures, with decreasing lattice spacing $a$, but
  keeping the volume constant, according to Table~2. The black line is the continuum extrapolation.
Right: an example of the continuum extrapolation  at a single value of the temperature $T=143\,$GeV.}\label{fig:phi_varb}
\end{figure}

For $m_H=115$\,GeV we measured $\Phi^\dagger\Phi$ at selected
temperature values while varying the lattice spacing by more
than a factor of two ($\beta_G = 6 \ldots 16$).  The results are shown
in Figure~\ref{fig:phi_varb}.  In this case we can reliably
take the continuum limit by linear extrapolation.  We observe that
when $\langle\phi^\dagger\phi\rangle$ is small, the cut-off effects are very
small, and at lowest temperatures $T \approx 130$\,GeV the $\beta_G=9$
result deviates from the continuum limit by less than 8\%. The physical volume
was kept fixed, at $L g_3^2 \approx 14$.  We have checked that this is a
large-enough volume so that the residual finite-volume effect is
unobservable within our statistical accuracy.

\subsection{The sphaleron rate $\Gamma_{\rm diff}(T)$}
\label{sec:resultsrate}

\begin{figure}
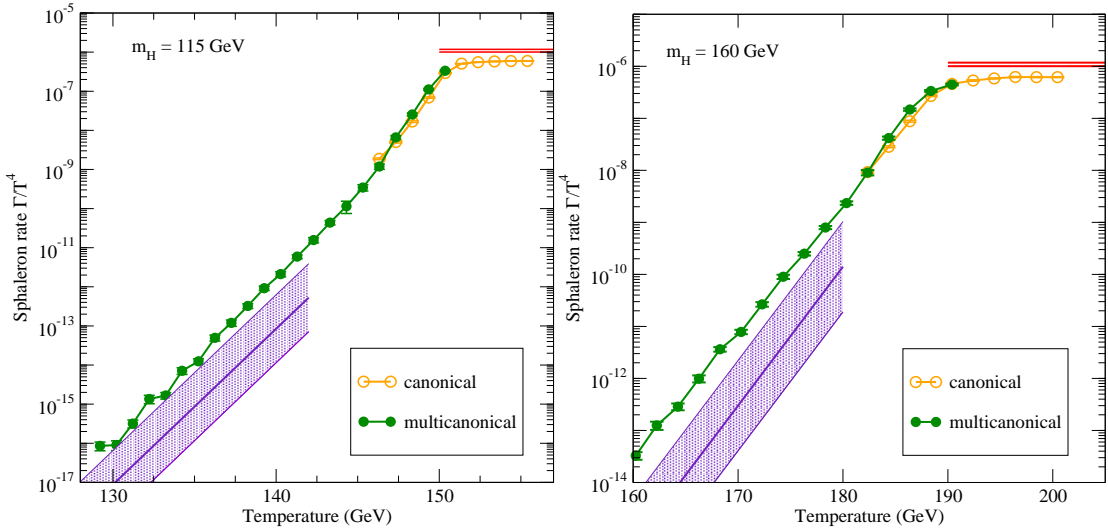

\begin{center}
\includegraphics[width=0.48\textwidth]{rate115naive.eps}%
\includegraphics[width=0.48\textwidth]{rate160naive.eps}
\end{center}
\caption[a]{The sphaleron rate for $m_H =115\,$GeV (left) and $160$\,GeV (right) at $\beta_G=9$. The shaded band is the
 theoretical estimate (plus error ranges extrapolated from lattice
 results in~\cite{Moore:1998swa}) for the broken phase and the horizontal lines for
 the symmetric phase, as calculated in~\cite{Burnier:2005hp}. 
The canonical and multicanonical results agree within errors at high temperatures.}
\label{fig:rates}
\end{figure}

The sphaleron rate $\Gamma/T^4$ for Higgs masses $115\,$GeV and $160\,$GeV is shown in Fig.~\ref{fig:rates}, using $\beta_G=9$ data. 
As expected, at high temperature in the symmetric phase, the rate becomes insensitive to the temperature apart from the trivial scaling. In this region the standard ``canonical'' real-time evolution is sufficient.  As we proceed into the cross-over region there is a rather sharp turnover, with a drop of $10^{-3}$ over $5\,$GeV. In this region, both the canonical and multicanonical methods were used, and they agree very convincingly. Deeper in the broken phase, the decrease in the rate flattens out somewhat to a clean exponential drop-off, and, using the multicanonical approach, we were able to follow the rate over 10 orders of magnitude. For comparison, we have included the extrapolation from~\cite{Burnier:2005hp}, expected to be valid deep in the broken phase. We see that the slope is correct, but that the central value of the rate is off by about an order of magnitude; or equivalently that the temperature axis is shifted by about $2\,$GeV for a Higgs mass of $115\,$GeV and about $5\,$GeV in the $160\,$GeV case.  
The data is shown for $\beta_G=9$, where the majority of our simulations
were done.

Sphalerons are extended objects, and thus it is necessary to check the finite
volume effects.  Using constant lattice spacing $\beta_G=9$ and lattice sizes $L/a = 16$--$54$ we observed no systematic finite-size dependence within our statistical
accuracy.   Thus, we
can be confident that $L = 32\,a \approx 14/g_3^2$ is safely large enough
at all temperatures.  This result is in agreement with 
ref.~\cite{Moore:1999fs}, where the volume dependence of the sphaleron rate became negligible at $L\, \gsim\, 5/g_3^2$ in pure SU(2) gauge theory.

\begin{figure}
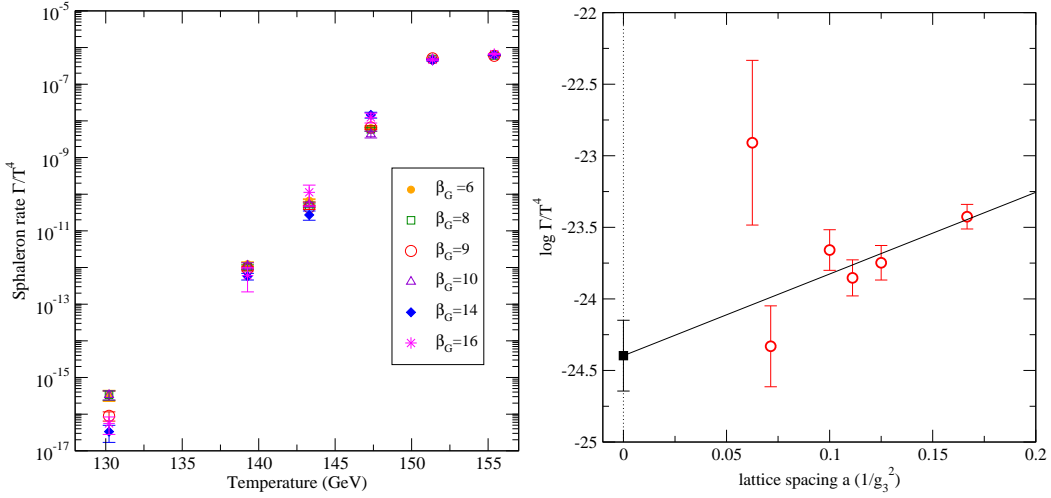

\begin{center}
\includegraphics[width=0.45\textwidth]{rate_fixedratio_naive.eps}
\includegraphics[width=0.45\textwidth]{log_gamma_extrap.eps}
\end{center}
\caption[a]{Left: the sphaleron rate calculated for $m_H =115\,$GeV and
  with different lattice spacings $a$ while keeping the physical
  volume constant. 
  Right: an example of the rate at  $T=143\,$GeV as a function of the lattice spacing.  Here the continuum limit has been extrapolated by assuming 
that the dominant error of $\log \Gamma$ is linear in $a$. 
However, if we extrapolate $\Gamma$ itself 
linearly in $a$ the result is compatible with vanishing rate in 
the continuum limit.}
 \label{fig:gamma_varb}
\end{figure}

As we did with the Higgs field expectation value, we investigated the dependence of the sphaleron rate on the lattice spacing.
We chose a set of six temperatures in the interval 130 -- 155\,GeV
and measured the rate at $\beta_G = 4/(g_3^2 a) = 6$--$16$, while keeping the physical volume approximately constant: $L \approx 3.5 \beta_G = 14/g_3^2$.
The lattice spacings and volumes are shown in Table~\ref{tab:bV},
and the resulting sphaleron rates are shown in Figure~\ref{fig:gamma_varb}.

In the symmetric phase the lattice-spacing dependence is very mild.
Deep in the broken phase the rate appears to decrease as $a$ is
decreased.  This can be understood in the light of the increasing
Higgs field expectation value at smaller $a$, see Figure~\ref{fig:phi_varb}.  

However, deep in the broken phase ($T\,\lsim\,
145$\,GeV) and for the smallest lattice spacings ($\beta_G \ge 14$) our
multicanonical order parameter, cooled $N_{\rm CS}$ (Figure~\ref{fig:cooling}), becomes ineffective and we are not able to obtain a
sufficiently accurate measurement of the rate for the proper continuum limit.
This is due to the increased noise in the measurement at smaller
lattice spacings: when the amplitude of the noise is of order unity,
a large fraction of the configurations with (measured) $N_{\rm CS}$ near
half-integer value are actually some distance from the top of the tunneling
barrier.  Thus, only a small fraction of these configurations will lead to tunneling trajectories.

The amount of noise  can be reduced by applying more cooling before the measurement of $N_{\rm CS}$.  However, cooling evolves the configuration towards one of the vacua ($N_{\rm CS}$ integer), and with too much cooling the measured order parameter does not track the true Chern-Simons number well enough for effective update.  We emphasize that despite these issues the
multicanonical method remains exact in the limit of infinite statistics;
it is only the efficiency of the method which suffers.  

Because of this issue our statistics at small lattice spacing is severely
restricted and we cannot obtain a reliable continuum limit.  Thus, our
final answer remains the $\beta_G=9$ result, where we have most of the data.
However, what the data indicates is that the true continuum limit is probably
a factor of 2--3 below the $\beta_G=9$ result deep in the broken phase, 
which very likely makes the agreement with
ref.~\cite{Burnier:2005hp} in Figure~\ref{fig:rates} better.

\begin{table}[!ht]
\begin{center}
\begin{tabular}[a]{|  p{2cm} |  p{2cm} |  p{2cm} |  p{2cm} |}
\hline

			$a$ ($1/g_3^2$) & $\beta_G$ & $L$ & $L / \beta_G$\\

			\hline

			0.67 & 6 & 20 & 3.3\\
			0.5 & 8 & 28 & 3.5\\
			0.44 & 9 & 32 & 3.56\\
			0.4 & 10 & 36 & 3.6\\
			0.29 & 14 & 48 & 3.4\\
			0.25 & 16 & 56 & 3.5\\

			\hline

\end{tabular}

\end{center}

\caption[a]{Lattice values for the continuum limit. $a$ is the lattice spacing, $\beta_G$ is defined in~(\ref{KLRS2.5}), $L$ is the size of our volume. From the ratio $L / \beta_G$ we notice that we keep the physical proportions constant while we diminish the size of the lattice spacing $a$.}
\label{tab:bV}

\end{table}

\section{A sample Leptogenesis calculation\label{sec:lepto}}

\begin{figure}
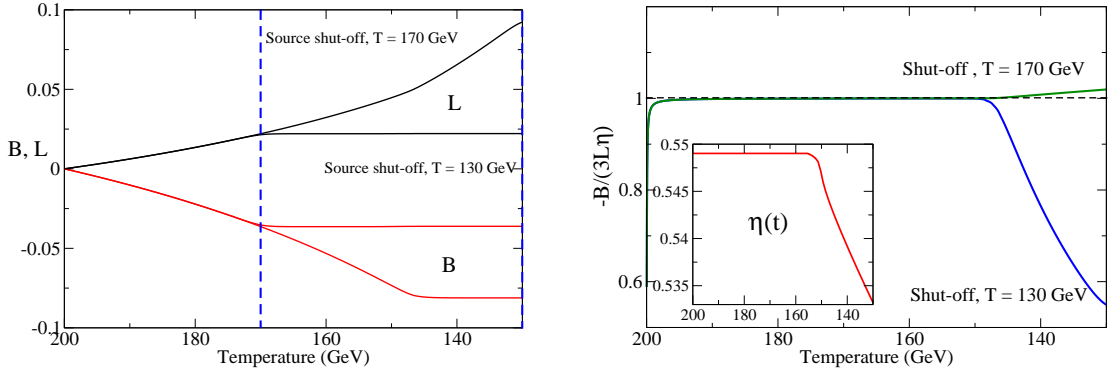

\vspace{4mm}
\begin{center}
\includegraphics[width=0.45\textwidth]{lepto1.eps}
~~~~~
\includegraphics[width=0.45\textwidth]{lepto2.eps}
\end{center}
\caption[a]{Left: the evolution of
  baryon number $B$ (red) and lepton number $L$ (black) in the
  presence of a lepton number source, turned on at $T=200\,$GeV. The
  source is turned off at $T\,=~170\,$GeV and $130\,$GeV respectively
  (blue dashed lines). 
  Right: The evolution of the
  ratio~(\ref{eq:equil}), with the source turned off at $T=170~$GeV
  (green) and $130\,$GeV (blue). Inserted: the evolution of the quantity
  $\eta(t)$. \label{fig:lepto}}
\end{figure}

To gauge the impact of using the
correct sphaleron rate and Higgs expectation value through the transition, we
solved the Leptogenesis equations~(\ref{eq:lep1})-(\ref{eq:lep2}), under
some simplifying assumptions. 

As the sphaleron rate is well-known in the symmetric phase, we focus our interest on the cross-over region, to investigate the efficiency of lepton-to-baryon number conversion through the newly-calculated sphaleron rate. In order to enhance the effect of the sphaleron rate suppression at cross-over temperatures, we study two limiting cases: one where the lepton-number source $f_i(t)$ was shut off well before the start of the cross-over, and one where we let the source active throughout.

We took $n_G=3$ and assumed that all lepton species are equivalent $L_i=L$,
  $i=1,2,3$, $\sum_i L_i=3L$. The initial baryon and lepton numbers
  vanish $L(t=0)=B(t=0)=0$. The source for the lepton number is therefore taken to
  be operational between $T=200\,$GeV and $T=T_{\rm cut-off}$, 
\begin{equation}
  f_i(t)=f(t)=\frac{f_0}{2}\left(1-\tanh\left[\frac{T_{\rm
          cut-off}-T}{2\,\textrm{GeV}}\right]\right),
\end{equation}
Since the equations are linear, the normalization of $f_0$ is
arbitrary. This leaves only the source shut-off temperature $T_{\rm
  cut-off}$ as a free parameter, which we varied 
  from
$170$ to $130\,$GeV\footnote{We chose the $2\,$GeV width to mimic a fast
  shut-off of the source.}, before and after the cross-over, respectively.

The full expressions for $\gamma(t)$ and $\eta(t)$ in~(\ref{eq:lep1}-\ref{eq:lep2}) read~\cite{Burnier:2005hp}
\begin{eqnarray}
\label{eq:leptoeq}
\gamma(t)=n_G^2\,\rho\left(\frac{v(T)}{T}\right)\left[1-\chi\left(\frac{v(T)}{T}\right)\right]\frac{\Gamma_{\rm diff}(T)}{T^3},\qquad \eta(t)=\frac{\chi\left(\frac{v(T)}{T}\right)}{1-\chi\left(\frac{v(T)}{T}\right)},
\end{eqnarray}
where $T=T(t)$, and the functions
\begin{eqnarray}
\label{eq:lep3}
\rho(x)&=&\frac{3\left[65+136n_G+44n_G^2+(117+72n_G)x^2\right]}{2n_G\left[30+62n_G+20n_G^2+(54+33n_G)x^2\right]},\\
\notag \\
\label{eq:lep4}
\chi(x)&=&\frac{4\left[5+12n_G+4n_G^2+(9+6n_G)x^2\right]}{65+136n_G+44n_G^2+(117+72n_G)x^2}.
\end{eqnarray}

We calculated the evolution of lepton and baryon number from
  temperature $200\,$GeV down to $130\,$GeV. Using that to a good
  approximation in the early Universe $T\propto 1/a$, with $a(t)$ the
  scale factor, we have that 
\be \frac{d}{dt}=-HT\frac{d}{dT},
\end{eqnarray}
where $H$ is the Hubble rate, given by the (radiation-dominated)
Friedman equation\footnote{We ignored the effect of $g^*$ changing
  slightly as the top quark begins acquiring its mass.}  
\be
H^2=\frac{\pi^2g^*T^4}{90 M^2_{\rm pl}},\quad g^*=106.75,\quad M_{\rm
  pl}=2.43\times 10^{18}\,\textrm{GeV}.
\end{eqnarray}
Over the range of temperatures used here, $H=(8.8-3.7)\times 10^{-14}$.

Once the source is turned off, and in the limit that $L$ and $B$
evolve much faster than $v$ and $\eta$, we can write the equations in
terms of $Y=B+\eta 3L$ \be \frac{d\ln Y}{d\ln
  T}=\frac{\gamma(T)}{H(T)}\left(1+\eta(t)\right),
\end{eqnarray}
so that $Y=0$ is enforced unless $\gamma/H$ is too small. And so if
the sphaleron rate is fast enough, we expect \be
\label{eq:equil}
-\frac{B(T)}{3L(T)\eta(T)}\simeq 1.
\end{eqnarray}
We say that the system is in ``equilibrium'' when this relation is
obeyed. We note that $\eta(v(T)/T=0)=0.549...$ and
$\eta(v(T)/T=\infty)=0.48$.

The evolution of $B$ and $L$ in
time is shown in Fig.~\ref{fig:lepto}~(left). Starting from zero at $T=~200\,$GeV, the introduction of the
source leads to a growing $L$ and, through sphaleron processes,
growing $B$. This continues until the source is switched off; in the examples
shown here $T_{\rm cut-off}=170\,$GeV and $T_{\rm cut-off}=130\,$GeV. 
For the early cut-off, both $B$ and $L$
level off to some asymptotic value. But even without switching off the
lepton source, at a temperature around $143\,$GeV the sphaleron rate
becomes inefficient, and the baryon number levels off. Lepton number
is still sourced, but having no longer $B$ as a sink, the growth of
$L$ becomes steeper. $T_{\rm freeze-out}$~$=$~$143$\,GeV corresponds to
$\gamma(t)/H\simeq 10$, and $v(T)/T\simeq 0.5$.

Fig.~\ref{fig:lepto}\,(right) shows the ``equilibrium'' condition~(\ref{eq:equil}) in time. We see an
initial transient, but from then on the system is nicely in
equilibrium until the freeze-out at $T=143\,$GeV. In this case, where
the source is still on (blue line), the equilibrium condition is
obviously broken by the continued sourcing of $L$. Since $L$ increases
and $B$ is constant, the ratio decreases.

What is perhaps more surprising is the ratio~(\ref{eq:equil}) when
there is no source. Because the $v(t)$ increases through the
transition, the equilibrium value $\eta(t)$ also changes (see
inset). A large enough sphaleron rate relative to the Hubble rate would adjust $B$ relative to $L$ to accommodate this evolving ``equilibrium'', but as is clear from
Fig.~\ref{fig:lepto}, this does not happen. $\eta(t)$ just decreases
and so the ratio becomes larger than $1$. As a consequence, the
asymptotic $B$ and $L$ obey
\begin{eqnarray}
B(x=\infty)\simeq \eta(T_{\rm freeze-out})3L(T_{\rm freeze-out})= 1.06\times 3L(T_{\rm freeze-out})\eta(x=\infty).
\end{eqnarray}

Fig.~\ref{fig:lepto} is based on the
$m_H=115\,$GeV data. We did a similar calculation for the
$m_H=160\,$ rate, giving the same picture but with $T_{\rm
  freeze-out}=175\,$GeV.

\section{Conclusion\label{sec:conclusion}}

In this paper we have presented the quantitative, non-perturbatively calculated sphaleron rate in the minimal Standard Model, using values of the Higgs mass still (marginally) allowed by experiment, $115\,$GeV and $160\,$GeV. At these Higgs masses, the electroweak transition is known to be an equilibrium cross-over.

We first probed the temperatures in order to find the range at which the cross-over takes place. We performed lattice simulations in the symmetric phase with canonical Monte Carlo methods using a straightforward heat-bath update algorithm of the Higgs and gauge fields. Here we found that the rate is unsuppressed and follows a random walk in time. When lowering the temperature to reach the broken phase, the rate becomes suppressed and in order to observe transitions between vacua we have to switch to multicanonical Monte Carlo and real-time simulations. 

We see a perfect match between the data points obtained with canonical and multicanonical Monte Carlo, with a smooth overlap in the cross-over region. This gives us a cross-check on the validity of the results. And by varying the physical volume and the lattice spacing, we found that both the sphaleron rate and Higgs field are close to their infinite-volume and continuum limits.

We obtained the sphaleron rate as a function of the temperature, showing an obvious similarity in the rate at the two Higgs masses $m_H=115\,$GeV and $160\,$GeV, Figure~\ref{fig:rates}. In both cases, the rates become exponentially suppressed in the broken phase with similar slopes, while having the same ($\sim 5 \times 10^{-7}$ T$^4$) asymptotic value in the symmetric phase. We point out that the cross-over range is clearly related to the Higgs mass: $130-147\,$GeV for $m_H =115\,$GeV, and $160 - 182\,$GeV for $m_H =160\,$GeV. The cross-over range is clearly noticeable also in the plots for the Higgs field.

Moreover, right at the beginning of the cross-over, the sphaleron rate drops at its fastest. Both in the plots for the sphaleron rate and for the Higgs field, the curves become steepest promptly after the cross-over kicks in, around $145-148\,$GeV for the smaller Higgs mass and $177-185\,$GeV for the bigger one.

The behaviour of the sphaleron rate we found is in agreement with results quoted in the literature, in the range where they exist. Our asymptotic value is of the same order of magnitude as in~\cite{Burnier:2005hp}. Direct comparison in the broken phase shows that while the slope is the same, the value is off by an order of magnitude.

Finally, we input the obtained sphaleron rate into a simple-minded model of Leptogenesis. We found that although the rate cuts off exponentially at the transition, the freeze-out of the baryon and lepton number happens slightly later, when the rate is about 10 in units of the Hubble rate. For precision calculations of the generated baryon asymmetry in such models, it is therefore important to take the gradual shut-off of sphaleron processes into account.

\acknowledgments This work is supported by the Academy of Finland
grants 114371 and 1134018.  The computations have been made at the
Finnish IT Center for Science (CSC), Espoo, Finland. M.\,D.~acknowledges support from the Magnus Ehrnrooth foundation. A.\,T.~is supported by the Carlsberg Foundation.

\end{document}